\documentclass{article}
\usepackage{graphicx}
\usepackage{color}

\title{Evidence for moving breathers in a layered crystal insulator at 300K}

\author{F. M. Russell, J. C. Eilbeck\\Department of Mathematics and
  the\\ Maxwell Institute for Mathematical Sciences\\ Heriot-Watt
  University \\Riccarton, Edinburgh, EH14 4AS, UK }

\date{\today}

\begin{document}

\maketitle
\begin{abstract}
  We report the ejection of atoms at a crystal surface caused by
  energetic breathers which have travelled more than $10^7$ unit cells
  in atomic chain directions.  The breathers were created by
  bombardment of a crystal face with heavy ions. This effect was
  observed at 300K in the layered crystal muscovite, which has linear
  chains of atoms for which the surrounding lattice has $C_2$
  symmetry.  The experimental techniques described could be used to
  study breathers in other materials and configurations.
\end{abstract}

The transport of energy through crystals has been studied extensively
in metals, but less so in insulators, in which the usually strong
electron-phonon coupling pathway for dispersion of energy is absent.
This absence allows long-range transport phenomena involving large
amplitudes of motion of atoms about their equilibriums positions to be
studied.  Large amplitudes of motion are created when a crystal is
irradiated by bombardment with energetic ions. Under irradiation some
material is sputtered backwards from the surface but most of the
incident energy is deposited in the body of the crystal.  If the
incident energy is sufficiently high, a plasma is created and leads to
the development of atomic cascades in which atoms are displaced
permanently from their equilibrium positions in the lattice.  This
stage in the coupling of energy to the lattice can be studied by
molecular dynamics (MD), which shows the emergence and distribution of
permanently displaced atoms \cite{kpgwsh01,skb97,nkga99}. A notable
feature of such cascades is the creation of focused collision
sequences, often terminating in an interstitial \cite{si57,vd97}.
These sequences are unstable against thermal motion because they
depend on the alignment of atoms \cite{ntm62}.  Typically, at ambient
temperatures, such sequences are limited to a length of about ten unit
cells. Eventually, as the incident energy is dispersed, atomic
displacements are no longer possible. It is at this stage, when the
available kinetic energy still far exceeds that of phonons, that
on-site potentials and long range co-operative interactions between
atoms can influence the subsequent dispersal of energy in the lattice
by the creation of self-focusing breathers \cite{rc95b,fw98}.  Here we
present evidence for the existence of energetic, mobile, highly
localised, lattice excitations that propagate great distances in
atomic-chain directions in crystals of muscovite, an insulating solid
with a layered crystal structure.  Specifically, when a crystal of
muscovite was bombarded at a given point, atoms were ejected from
remote points on another face of the crystal, lying in atomic chain
directions and at more than $10^7$ unit cells distance from the site
of bombardment. This effect was predicted from studies of inelastic
scattering of breathers using analogue mechanical-magnetic arrays and
has been confirmed by numerical simulations.  As this experiment was
performed with the crystal at around 300K, it demonstrates the
stability of these breathers against thermal motions of the lattice.

The evolution of an impulse to a single particle in a discrete
particle system consisting of a chain of particles with nonlinear
nearest neighbour interactions has been studied extensively and leads
to solitons \cite{to67}.  However, in two-dimensional arrays of such
chains these solitons are unstable, limiting flight paths to about 100
unit cells \cite{mm91} In real crystals the evolution of an impulse is
strongly influenced by the surrounding lattice, which introduces
on-site potentials. Both mechanical analogues and numerical methods
have been used to study how an impulse to a single atom in a chain
evolves under these conditions, in particular, for the muscovite
lattice. On a time scale of a few picoseconds, it evolves via
anharmonic oscillations into a highly localised, mobile, excitation in
which adjacent atoms in a chain move in nearly anti-phase motion
within an envelope extending over a small ($\sim$10) number of atoms.
The envelope also extends in a lateral direction, but most of the
energy resides on a single chain. These excitations are a type
localization called variously an Intrinsic Localised Mode or a
breather \cite{fw98,shs06}.  First order numerical modelling of the
mica lattice showed that these breathers could propagate up to about
$10^4$ unit cells before becoming unstable \cite{mer98}.

\begin{figure}[htb] %fig 0
 \begin{center}   \includegraphics[scale=0.6]{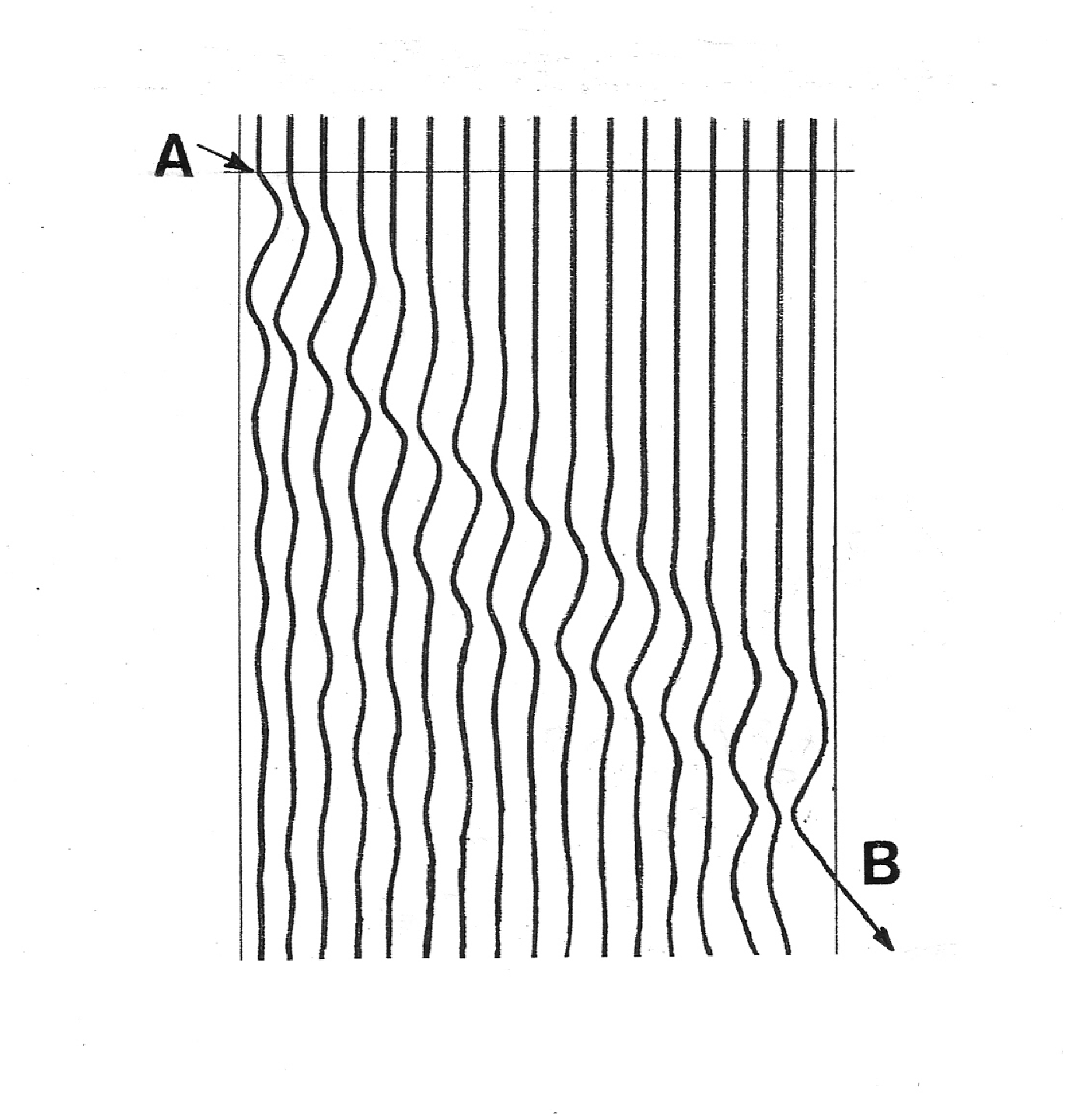} 
   \caption{Analogue plot of the longitudinal motion of particles
     lying in a chain when the first particle is given an impulse at
     point $A$. The impulse evolves into a breather that propagates
     along the chain.  Inelastic scattering of the breather when it
     reaches the end of the chain causes the last particle to be
     ejected over a potential energy barrier, as shown at point $B$.}
\label{fig0}\end{center}
\end{figure}

The requirement that atoms do not become interstitial sets an upper
limit on the energy of individual atoms in a breather of $\sim$50 eV.
This estimate is derived from plots of potential energy versus
displacement, calculated for the muscovite lattice by molecular
dynamic methods.  A mechanical analogue using magnetic dipoles was
constructed, which mimicked the potential energy versus displacement
plot. This analogue showed the evolution of an initial impulse into a moving
breather, its behaviour on meeting defects in the chain, such as
different masses or a vacancy, and reflection at the end of a chain.
These studies indicated that breathers could have energies from a few eV
up to $\sim$100 eV.  Further, it showed that breathers with more than
$\sim$10 eV could suffer inelastic scattering at the end of a chain
and eject the last atom. This process is illustrated in Fig.\
\ref{fig0}. The prediction that energetic breathers could eject atoms
is testable and was the rationale for the experiment reported here.

\begin{figure}[ht] %fig 1
 \begin{center}   \includegraphics[scale=0.6]{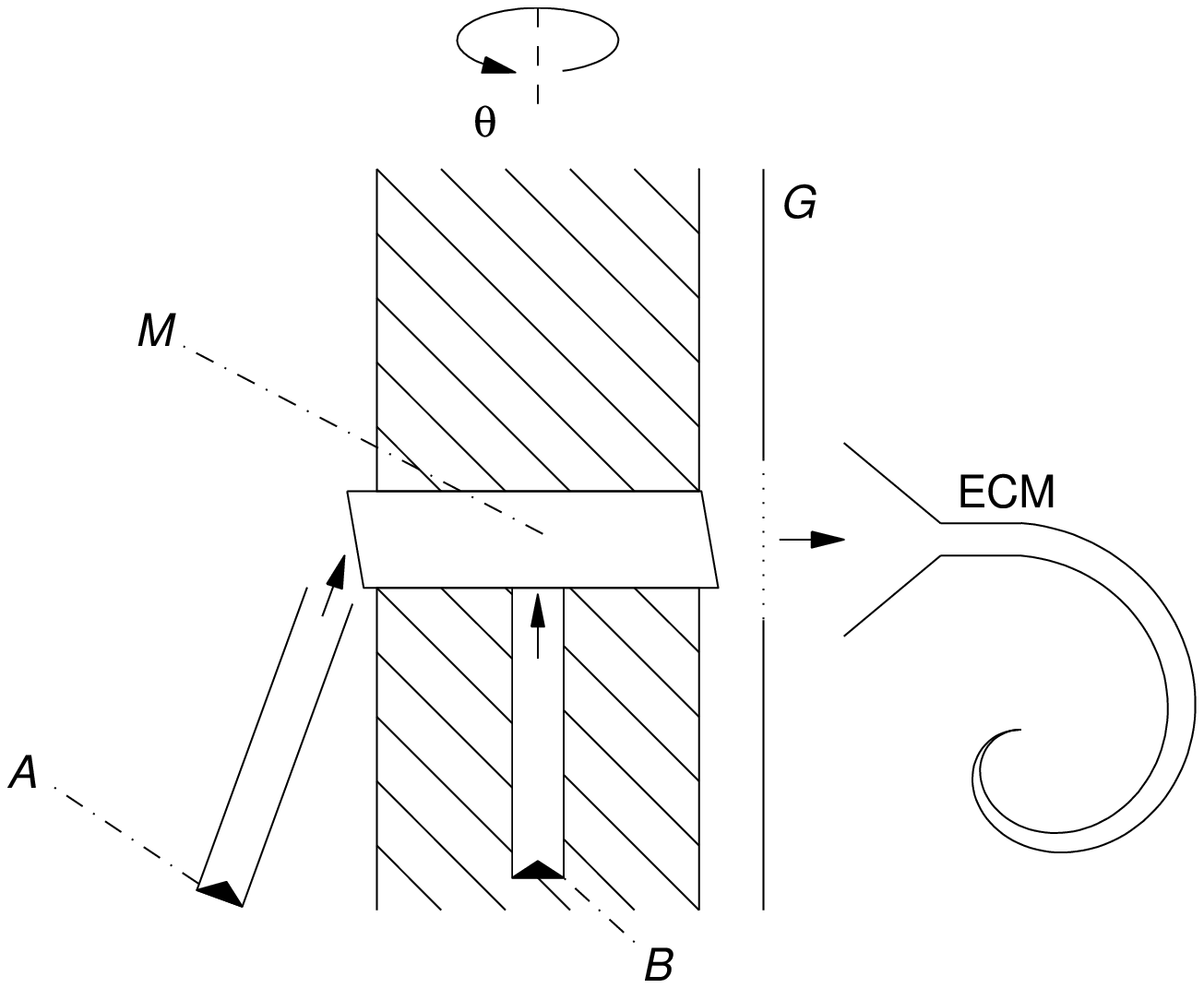} 
    \caption{Schematic diagram of the experiment. The crystal of
      muscovite $M$ was clamped between metal plates pressing on the
      (001)-faces. In the first arrangement $A$, the collimated beam of
      $\alpha$-particles bombarded one edge. Atoms ejected from the
      opposite edge, after ionization, passed through the grid $G$ and
      were detected by the electron channel multiplier ECM.  In the
      second arrangement $B$, the source is placed below the crystal.}
\label{fig1}\end{center}
\end{figure}

\begin{figure}[ht] %fig 2
\begin{center}
    \includegraphics[scale=0.6]{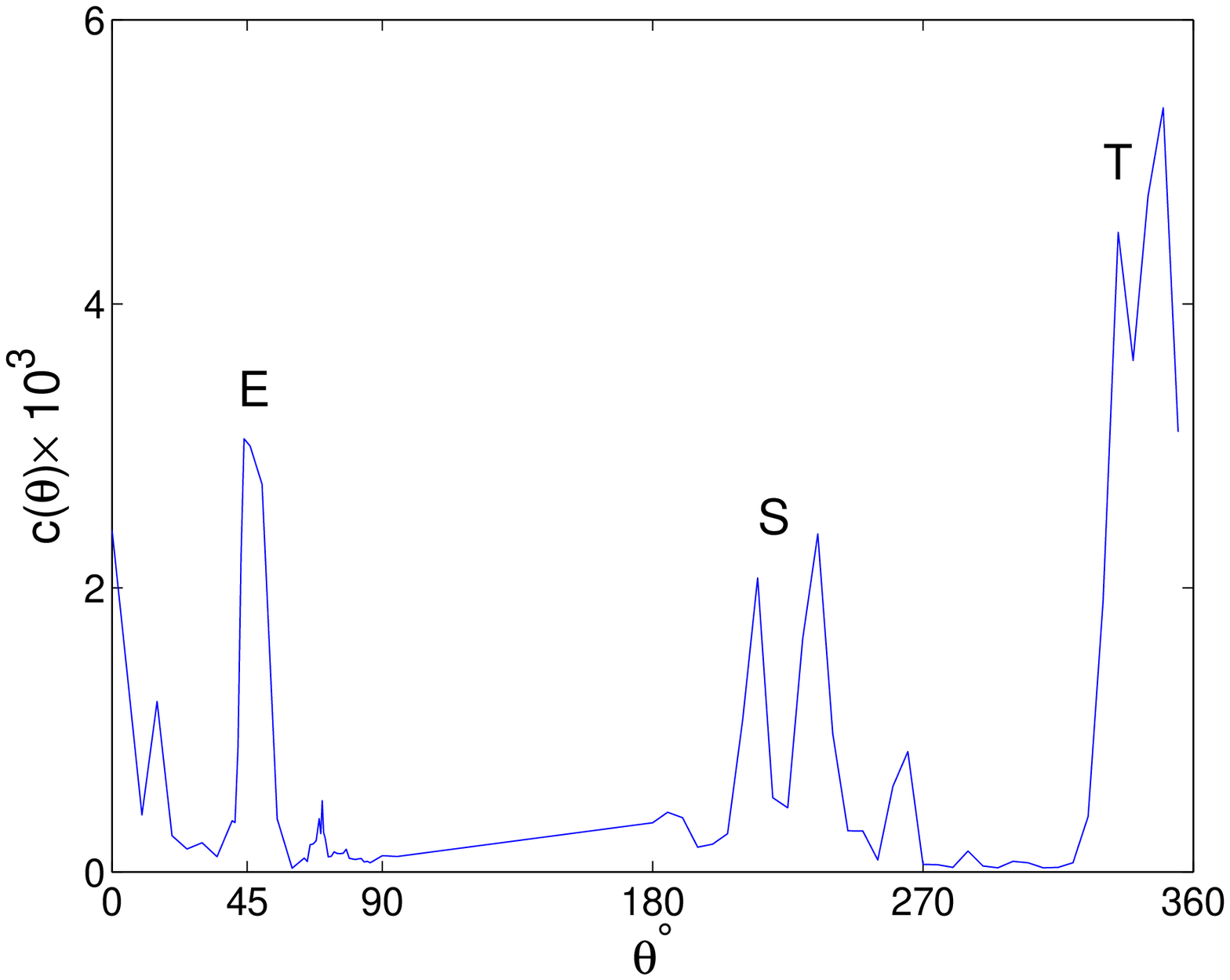} 
    \caption{Plot of the angular dependence of the ECM count rate. The
      peak labelled $T$ arises from a source of alpha particles used
      for test purposes.  The several peaks $S$ arise from both back
      scattering and sputtering from the front face, when it is
      irradiated with alpha particles. Breathers are created during
      the irradiation and propagate in the [010] chain direction. On
      reaching the rear face of the crystal, after travelling more
      than $10^7$ unit cells through the lattice, those breathers with
      sufficient energy eject atoms giving rise to the peak labelled
      $E$. As the crystal was held at about 300${}^\circ$K, this result
      demonstrates the stability of breathers against thermal motion,
      in contrast to the short range of classical collision cascades.}
\label{fig2}
\end{center}
\end{figure}

The experiment is shown schematically in Fig.\ \ref{fig1}. 
It consisted of an electrically insulated rotatable specimen holder
placed at the centre of a vacuum chamber and positioned opposite an
exit port to which a Photonis X919AL electron channel-plate multiplier
[ECM] was attached. This exit port was covered by a fine wire grid to
help shape the electric field between the specimen and the ECM when a
positive potential of several kV was applied to the specimen holder. A
small crystal of muscovite of $\sim$1mm thickness and $\sim$7mm across
the (001)-face with good natural edge faces was mounted on the
specimen holder with the axis of rotation of the holder normal to the
(001)-face. One edge of the crystal was bombarded with
$\alpha$-particles of $\sim$5Mev from an Am${}^{241}$ source $A$
located at the bottom of a collimating tube, which also shielded the
ECM from the $\alpha$-particles. This tube was positioned to irradiate
one edge of the crystal at near grazing incidence and approximately
normal to the (001)-face. A positive potential was applied to the
crystal holder to create an electric field approximately normal to the
edges. Atoms ejected from the  crystal edges and ionized by
electrons moving in this electric field were accelerated towards the
ECM for detection. The count rate of the ECM was measured as the
crystal holder was rotated. The result is shown in Fig.\ \ref{fig2}.

The peak labelled $T$ is caused by a source of alpha particles mounted
on the target holder but screened from the crystal for test and
calibration purposes of the detection system. The several peaks
labelled $S$ are due to sputtering and back scattering from the
irradiated face. Here we are mainly interested in the peak labelled
$E$ which is due to atoms ejected from the rear face by inelastic
scattering of breathers. The angular position of this peak is
consistent with ejection from chains lying in the principal (010)
crystal direction, after the breathers had travelled a distance of
$10^7$ unit cells.

To eliminate the possibility of spurious secondary particles arising
from the irradiated front face and then multiply scattered via the
chamber to the back face, the reference source was removed and the
source used to irradiate the crystal was placed below the crystal in a
holder that collimated the alpha-beam, as indicated in Fig.\
\ref{fig1} as source $B$. The source holder could be rotated to bring
the source under a hole in the base of the target holder, so that the
flux of alphas could be controlled. It was found that the ECM count
rate for the peak $E$ was proportional to the alpha flux, as expected.

The absence of a signal from the ECM at zero applied potential showed
that the peak $E$ was not due to $\alpha$-particles channelling through the
crystal; the expected range of the alphas in the (001) plane from the
source $A$ was less than 10 microns. It was found that the flux of
ejected particles decreased slowly in time, with other variables held
constant, consistent with a progressive depletion of atoms that could
be ejected by inelastic scattering of the breathers. The decay was
approximately exponential in time, with a decay constant of $\sim$4
hours. Possible alternative sources of the ions detected by the ECM
were examined. The drop to background count rate either side of the
peak $E$ showed that the ions were not arising from out-gassing of the
crystal. Also, when a crystal with a freshly cleaved edge was used, it
gave similar results to that for a natural edge, implying that the
peak $E$ was not arising from surface contamination. This demonstration
that breathers can propagate great distances also shows that they can
survive the inevitable point defects that occur in real crystals as
well as remaining stable in the presence of considerable thermal
motion at $\sim$300K. Having remained stable whilst propagating more than
$10^7$ unit cells, there is no a priori reason to suppose that such
breathers could not travel much further in large nearly perfect
crystals.

We remark that these results constitute a useful investigative tool
for studying the formation and properties of breathers in a wide
variety of materials and material configurations.  We note also that,
although breathers in metals may not have such long path lengths, they
may still be important in various effects which are still poorly
understood using conventional theories.  We conclude by speculating on
two possible developments in this area.

Although these results relate to layered crystals, there is evidence
that breathers can occur in non-layered crystals, but with shorter path
lengths of order microns. This was reported in connection with
radiation damage studies in silicon \cite{sar00} and in diffusion of
interstitial ions in austenitic stainless steel \cite{am06}. It has
also been shown that the lattice conditions required for the existence
of breathers, namely linear chains of atoms for which the surrounding
lattice has $C_2$ symmetry, also occur in typical high temperature
superconductors \cite{rc96,mre01}.  The demonstrated stability of
breathers against thermal motion and their ubiquitous occurrence,
leads us to predict that reducing breather-breather scattering in the
conductive sheets of these superconductors, by separating the sheets
into narrow parallel strips, should increase the $T_c$ beyond the
present values. This could form a test of the supposition that
breathers are involved in pair formation in these layered materials.

It has also been reported that deuterium fusion reactions can be increased
by implanting deuterium in a metal \cite{chbhhr01}.  This effect has
been explained in part as due to electron screening. Due to the
ubiquitous nature of breathers, they will be produced copiously in such
experiments. During the evolution of a breather from the initial impact,
it is likely that several close encounters occur for each $d-d$
pair, which should contribute to an increased fusion rate.  This
may be an alternative or an enhancement  mechanism for this effect.

\noindent {\bf Acknowledgements} We wish to thank Drs.\ L.\ Cruzeiro
and R.\ Webb for helpful discussions and Dr.\ A Pring, Museum of South
Australia, for supplying some of the mica crystals.

%\bibliography{re06}
%\bibliographystyle{unsrt}

\end{document}